\documentclass[twocolumn,showpacs,showkeys]{revtex4}
\usepackage{graphicx}
\usepackage{bm}
\usepackage{color}
\usepackage{amsmath}
\usepackage{natbib}

\begin{document}

\title{Langmuir waves in semi-relativistic spinless quantum plasmas}

\author{A. Yu. Ivanov}%
 \email{alexmax1989@mail.ru}

\author{P. A. Andreev}%
\email{andreevpa@physics.msu.ru}

\author{L. S. Kuzmenkov}%
 \email{lsk@phys.msu.ru}

\affiliation{%
Physics Faculty, Moscow State University, Moscow, Russian
Federation.}

 \date{\today}

\begin{abstract}
Many particle quantum
hydrodynamics based on the Darwin Hamiltonian (the Hamiltonian corresponding to the Darwin Lagrangian) is considered. A force field appearing in
corresponding Euler equation is considered in details.
Contributions from different terms of the Darwin Hamiltonian in the
Euler equation are traced. For example, the relativistic correction to the
kinetic energy of particles leads to several terms in the Euler
equation, these terms have different form. One of them has
a form similar to a term appearing from the Darwin term.
Hence, the two different mechanisms give analogous contributions in wave
dispersion. Microscopic analog of
the Biot-Savart law, called the current-current interaction and
describing an interaction of moving charges via the magnetic field, is
also included in our description. The semi-relativistic
generalization of the quantum Bohm potential is obtained.
Contribution of the relativistic effects in the spectrum of plasma
collective excitations is considered.
\end{abstract}


\maketitle


\section{Introduction}

There is fast growing interest to the theory of the relativistic \cite{Haas arxiv 12}-\cite{Asenjo PP 11} and the semi-relativistic (weakly-relativistic) \cite{Asenjo arxiv 2011} quantum plasmas.
In this paper we develop the many-particle quantum hydrodynamics (QHD) \cite{MaksimovTMP 1999}-\cite{Andreev arxiv 11 3} in the semi-relativistic
approximation. In this way
we are going to discuss relations between the quantum, thermal, and semi-relativistic
effects in the system of many charged particles. As the result, we present complete theory including effects mentioned above for the spinless charged particles.

Spin leads to effects appearing in the semi-relativistic
approximation. However, it plays significant role in
non-relativistic physical systems, for example, in
ferromagnetic materials. Spin dynamics is also very important in physics of the quantum plasma, where electrons and positrons are most widespread
objects, and their spin is an inherent dynamical property. Over the last
decade a lot of papers have been dedicated to studies of spin dynamics in
quantum plasmas, especially by means of the quantum hydrodynamics and
Vlasov-like kinetic equations. However, it is very interesting
and important to understand the quantum many-particle physics
appearing from consideration of the Darwin Hamiltonian the Hamiltonian corresponding to the Darwin Lagrangian, which is
the spinless analog of the Breit Hamiltonian \cite{Landau 4}. The Darwin
Hamiltonian contains both the non-relativistic terms, which
describe kinetic energy of particles and the Coulomb interaction, and
semi-relativistic terms. They describe the relativistic correction to the
kinetic energy of particle (RCKE), the interaction energy of moving
charges (which is also called the current-current interaction), the Darwin term proportional to $\nabla \textbf{E}$ and the term, describing interparticle interaction,
proportional to the Dirac delta function, corresponding to the Darwin term, we call it the Darwin interaction. The current-current interaction
presents the Biot-Savart-Laplace law. The RCKE and the current-current interaction should be important when studying the relativistic beams in the plasmas.

Suggestion was made in Ref. \cite{Asenjo arxiv 2011} that in
some cases contribution of the RCKE much smaller than the
Darwin term. However, our studies of the semi-relativistic
effects in the quantum plasmas based on the quantum hydrodynamics method
show that the RCKE leads to existence of terms in the
semi-relativistic Euler equation. One of these terms has form
close to the only term brought by the Darwin term and Darwin interaction. Thus the Darwin term and the RCKE must be considered together.

Contribution of the RCKE and the current-current interaction in the
plasma wave dispersion have been
considered recently \cite{Ivanov RPJ 13} in terms of the many-particle quantum hydrodynamics developed in Refs. \cite{MaksimovTMP 1999}-\cite{Andreev
arxiv 11 3}, but the Darwin term was not considered there. Another derivation of the QHD equations for
systems of charged spinning particles suggested later can be found in Refs.
\cite{Marklund PRL07}, \cite{Brodin NJP07}. Some aspects of the quantum
plasma physics were reviewed in Ref. \cite{Shukla RMP 11}.

The Darwin term is the semi-relativistic trace of the Zitterbewegung contribution in the Langmuir wave dispersion. The Zitterbewegung effect has been actively studied \cite{Lamata NJP 11}-\cite{Boada NJP 11}. It has been considered for electrons in semiconductors \cite{Zawadzki JP CM 11}, \cite{Biswas JP CM 12}, ions \cite{Lamata NJP 11}-\cite{Schneider RPP 12} and the quantum gases of neutral atoms \cite{Zhang FP 12}-\cite{Boada NJP 11}. Consequently it is worthwhile to point out that the RCKE gives the contribution in the equations of collective motion counteractive to the Darwin term and Darwin interaction contributions.

This paper is dedicated to comparison of the RCKE, the
Darwin interaction and the current-current interaction contributions in the Euler
equation, obtaining of the explicit form of the semi-relativistic pressure
tensor and its influence on the dispersion properties of the longitudinal waves.

Our paper is organized as follows. In Sec. II we discuss basic
Hamiltonian and compare contributions of different terms. In Sec.
III a set of QHD equations is presented in semi-relativistic approximation. Different contributions in the Euler equation are discussed. In Sec. IV the method of dispersion equation obtaining is
described, linearized set of the semi-relativistic Euler equations
is presented. In Sec. IV dispersion relation for the quantum
semi-relativistic Langmuir waves is calculated and discussed. In
Sec. V brief summary of obtained results is presented.

\section{The model description}

The equations of quantum hydrodynamics are derived from the
non-stationary Schrodinger equation for system of N particles:
\begin{equation}\label{SRW Schrodinger}\imath\hbar\partial_{t}\psi(R,t)=\hat{H}\psi(R,t)\end{equation}
with Hamiltonian
\begin{equation}\label{SRW Hamiltonian}\hat{H}=\hat{H}_{0}+\hat{H}_{Rel}+\hat{H}_{D},\end{equation}
where
\begin{equation}\label{SRW Hamiltonian non rel} \hat{H}_{0}=\sum_{i}\biggl(\frac{1}{2m_{i}}\textbf{D}_{i}^{2}+e_{i}\varphi_{i,ext}\biggr)+\frac{1}{2}\sum_{i,j\neq
i}e_{i}e_{j}G_{ij},\end{equation}
\begin{equation}\label{SRW Hamiltonian relativ} \hat{H}_{Rel}=-\sum_{i}\frac{1}{8m_{i}^{3}c^{2}}\textbf{D}_{i}^{4}-\frac{1}{2}\sum_{i,j\neq
i}\frac{e_{i}e_{j}}{2m_{i}m_{j}c^{2}}G^{\alpha\beta}_{ij}D_{i}^{\alpha}D_{j}^{\beta},\end{equation}
and
$$\hat{H}_{D}=-\sum_{i}\frac{e_{i}\hbar^{2}}{8m_{i}^{2}c^{2}}\nabla_{i}\textbf{E}_{i,ext}$$
\begin{equation}\label{SRW Hamiltonian Darwin} -\frac{1}{2}\sum_{i,j\neq
i}\frac{\pi e_{i}e_{j}\hbar^{2}}{2c^{2}}\biggl(\frac{1}{m_{i}^{2}}+\frac{1}{m_{j}^{2}}\biggr)\delta(\textbf{r}_{i}-\textbf{r}_{j}).\end{equation}
This Hamiltonian corresponds to the spin independent part of the Breit Hamiltonian (see \cite{Landau 4} sections 33 and 83, and \cite{Strange book} formula (4.74b)). All terms except the fourth and sixth terms also correspond to the classic Hamiltonian derived from the Darwin Lagrangian (see \cite{Landau 2} section 65). The
following designations are used in the equation (\ref{SRW Hamiltonian})-(\ref{SRW Hamiltonian Darwin}): $e_{i}$, $m_{i}$ are the charge and the mass of particle, $\hbar$ is the Planck constant and $c$ is the speed of light,
$D_{i}^{\alpha}=-\imath\hbar\partial_{i}^{\alpha}-e_{i}A_{i,ext}^{\alpha}/c$ is the covariant derivative,
$\varphi_{i,ext}$, $A_{i,ext}^{\alpha}$ the potentials of an
external electromagnetic field,
$\partial_{i}^{\alpha}=\nabla_{i}^{\alpha}$ is the spatial derivatives,
$G_{ij}=1/r_{ij}$ is the Green functions of the Coulomb
interaction, $\textbf{r}_{ij}=\textbf{r}_{i}-\textbf{r}_{j}$,
\begin{equation}\label{SRW CC Gr func} G^{\alpha\beta}_{ij}=\frac{\delta^{\alpha\beta}}{r_{ij}}+
\frac{r^{\alpha}_{ij}r^{\beta}_{ij}}{r_{ij}^3}\end{equation} is
the Green functions of the current-current interaction,
$\psi(R,t)$ is psi-function of N particle system,
$R=(\textbf{r}_{1},...,\textbf{r}_{N})$. Let us consider physical
meaning of the terms in the Hamiltonian (\ref{SRW Hamiltonian}).

We consider the Hamiltonian as the sum of three parts: the non-relativistic part $H_{0}$, the relativistic part $H_{Rel}$, and the quantum-relativistic terms $H_{D}$. The first term in the non-relativistic part of the Hamiltonian $H_{0}$ is the
kinetic energy, the first term in $H_{Rel}$ is the RCKE, the second term in $H_{0}$ is the potential
energy of the classic charge in the external electric field, the
first term in $H_{D}$ is the quantum contribution in the energy of the charge
being in the external electric field, which is called the Darwin
term. All these terms are valid for each particle, as they
describe kinematic properties and interaction with the external field. They present the first groups of terms in Hamiltonians (\ref{SRW Hamiltonian non rel}), (\ref{SRW Hamiltonian relativ}) and (\ref{SRW Hamiltonian Darwin}).
The second groups of terms in $H_{0}$, $H_{Rel}$, $H_{D}$ describe inter-particle interactions. First of all, the Coulomb interaction is presented by the third term in $H_{0}$. The second
term in $H_{D}$ describes a quantum contribution to the interaction of charges.
It is the Darwin interaction. The second term in $H_{Rel}$ describes the
current-current interaction, which is the microscopic analog of
the Biot-Savart law.

The
first term in $H_{D}$ shows a semi-relativistic contribution to the force
acting from the external electric field on charged particle (the Darwin term). The second
term presents the interaction between two particles, which
can be considered as a semi-relativistic addition to the Coulomb
interaction (the Darwin interaction). If we have deal with interaction of two electrons, the Darwin interaction is
\begin{equation}\label{SRW Ham Dar sim} H_{D}=-\pi\biggl(\frac{ e\hbar}{m c}\biggr)^{2}\delta(\textbf{r}_{i}-\textbf{r}_{j}),\end{equation}
where $\textbf{r}_{i}$ and $\textbf{r}_{j}$ are the coordinates of
the two electrons. The explicit form of the Darwin interaction
was derived from the scattering amplitude in the quantum electrodynamics. Now
we have to compare the Darwin term describing
interaction with the external field, the first term in formula
(\ref{SRW Hamiltonian relativ}), which appearing in the semi-relativistic
limit of the Dirac equation \cite{Landau 4}, and the Darwin interaction presented by the second term in $H_{D}$ \cite{Landau 4}.
Admitting that
$\triangle_{i}(1/|\textbf{r}_{i}-\textbf{r}_{j}|)=-4\pi\delta(\textbf{r}_{i}-\textbf{r}_{j})$
and introducing the microscopic electric field caused by particle $j$ acting on particle $i$ as
$\textbf{E}_{ij}=-\nabla_{i}(e_{j}/r_{ij})$, we see that the second term in
Hamiltonian (\ref{SRW Hamiltonian Darwin}) can be represented as
\begin{equation}\label{SRW Dar non sim} H_{D}=-\frac{ e_{i}\hbar^{2}}{8c^{2}}\biggl(\frac{1}{m_{i}^{2}}+\frac{1}{m_{j}^{2}}\biggr)\nabla_{i}\textbf{E}_{ij}.\end{equation}
In formula (\ref{SRW Dar non sim}) we used general dependence of masses for interacting particles obtained in Ref. \cite{Landau 4}. In formula (\ref{SRW Ham Dar sim}) we have assumed $m_{i}=m_{j}=m$. Comparing the first term in the Hamiltonian (\ref{SRW Hamiltonian Darwin})
and formula (\ref{SRW Dar non sim}) we get that these terms
coincide if $m_{j}\rightarrow\infty$ that corresponds to the Dirac equation. The Dirac equation describes motion of an electron in an external field, so motion of the electron gives no influence on the external field. Consequently, a mass of source of external field can be considered as equal to infinity. However, if we consider
interaction of two electrons, we have $m_{i}=m_{j}$ and from
(\ref{SRW Ham Dar sim}) we find
\begin{equation}\label{SRW}H_{D}=-\frac{ e\hbar^{2}}{4 m^{2}c^{2}}\nabla_{i}\textbf{E}_{ij},\end{equation}
what differs in two times from the first term in (\ref{SRW Hamiltonian Darwin}). It was expected that discussed terms should coincide due to the superposition principle, so we have to put additional factor two
in the first term in the Hamiltonian (\ref{SRW Hamiltonian Darwin}), but
we keep in mind that we can make another choice and accept
the consequence of the Dirac equation. At discussion of wave
dispersion we consider consequences of the both choices.

We have deal with the methods based on a certain equation, in our case it is the Schr\"{o}dinger equation (\ref{SRW Schrodinger}), describing system evolution in terms of the Hamiltonian for particles. In the relativistic case the Dirac equation is the corresponding equation. However the Dirac equation describes the quantum motion of one relativistic electron in an external electromagnetic field. There is no proper equation describing the quantum or classic motion of many relativistic electrons in terms of a Hamiltonian, since the Hamiltonian of electromagnetic field has to be included and the field should be considered as independent variable, as it is in the quantum electrodynamics. Thus there is no proper many-particle generalization of the Dirac equation. Consequently the Dirac equation do not allow to derive many-particle relativistic hydrodynamic directly. Even semi-relativistic hydrodynamics could not be derived by means the Dirac equation. However, the Breit Hamiltonian obtained from the quantum electrodynamic scattering amplitude of two charged spinning particles describes the semi-relativistic system of two particles (see Ref. \cite{Landau 4} section 83). It is easy to generalize the Breit Hamiltonian on system of $N$ particles, where $N>2$. Including the fact that we consider spinless particles, we see that the many particle Breit Hamiltonian corresponds to the classic Hamiltonian obtained from the Darwin Lagrangian (see Ref. \cite{Landau 2} section 65). But the Breit Hamiltonian contains the Darwin term and the Darwin interaction having quantum semi-relativistic nature. So it does not appear in the classic semi-relativistic theory.

For short
references below we introduce new function $\tilde{G}_{ij}$, which
is defined as
$\tilde{G}_{ij}=G_{ij}-(\hbar^{2}/4m^{2}c^{2})\delta(\textbf{r}_{i}-\textbf{r}_{j})$.

$\tilde{G}_{ij}$ leads to existence of two force
field terms in the Euler equation. Let us consider how they emerge during
derivation of the semi-relativistic Euler equation. We differentiate the current $\textbf{j}$ appearing in the continuity
equation with respect to time and use the Schr\"{o}dinger equation. One
of these terms appears due to commutation of $\tilde{G}_{ij}$ with the
momentum operator $\hat{p}_{i}^{\alpha}$ in the current
$\textbf{j}$. Let us point out that the operator
$\hat{p}_{i}^{\alpha}$ exists in the current $\textbf{j}$ due to presence of the kinetic energy
operator in the Hamiltonian (\ref{SRW
Hamiltonian}). In the self-consistent field approximation this
term has following form:
\begin{equation}\label{SRW Coulomb int form}\textbf{F}_{C}=-e^{2}n\nabla\int d\textbf{r}' \biggl(\frac{1}{\mid\textbf{r}-\textbf{r}'\mid}-\frac{\pi \hbar^{2}}{m^{2}c^{2}}\delta(\textbf{r}-\textbf{r}')\biggr)n(\textbf{r}',t).\end{equation}
The self-consistent field approximation allows to introduce the
electric field $\textbf{E}$ caused by the charges.
It has following explicit form
\begin{equation}\label{SRW }\textbf{E}=-e\nabla\int d\textbf{r}' \frac{1}{\mid\textbf{r}-\textbf{r}'\mid}n(\textbf{r}',t),\end{equation}
where $n$ is the particle concentration, and field \textbf{E} satisfy to the quasi-electrostatic Maxwell equations:
\begin{equation}\label{SRW Max rot}\nabla\times\textbf{E}(\textbf{r},t)=0,\end{equation}
and
\begin{equation}\label{SRW Max div}\nabla\textbf{E}(\textbf{r},t)=4\pi\sum_{a}e_{a}n_{a}(\textbf{r},t),\end{equation}
where subindex "a" describes species of particles. We are interested in dispersion of the Langmuir waves, which are the high frequency oscillations. Consequently electrons give main contribution there. Thus we can neglect motion of ions and consider ions as motionless. Having mixture of electrons and ions we work with a stable system. Presence of motionless ions reveals in the Poisson equation (\ref{SRW Max div}), where ions cancel equilibrium charge density of the electrons. So perturbations of electric field are caused by the perturbation of the electron density
\begin{equation}\label{SRW Max div one species}\nabla\textbf{E}(\textbf{r},t)=4\pi e \delta n(\textbf{r},t),\end{equation}
where $\delta n$ is the perturbation of electron concentration.

Now force field $\textbf{F}_{C}$ (\ref{SRW Coulomb int form}) takes form
\begin{equation}\label{SRW Coulomb wide concentr}\textbf{F}_{C}=en\textbf{E}+\frac{\pi e^{2}\hbar^{2}}{m^{2}c^{2}}n\nabla n,\end{equation}
where the concentration under the space derivative represents
source of the field. So, using equation
(\ref{SRW Max div one species}) for the mentioned concentration we have come to
\begin{equation}\label{SRW Coulomb wide field}\textbf{F}_{C}=en\textbf{E}+\frac{e\hbar^{2}}{4m^{2}c^{2}}n\nabla(\nabla\textbf{E}).\end{equation}
Presented here form of the second term corresponds to the
semi-relativistic contribution in the force acting on the charged
particle from external electric field obtained from the Dirac
equation \cite{Landau 4}. The second term in formula (\ref{SRW Coulomb wide concentr}) gives general form of the Darwin interaction force field. In some cases it can be rewritten in terms of the self-consistent electric field. We have done this representation for electron-ion plasmas with motionless ions. Formula (\ref{SRW Coulomb wide field}) is useful for comparison of the Darwin interaction with the RCKE.

The second term associated with $G_{ij}$ appears due to the RCKE. Or, more precisely, it exists due to simultaneous account of the RCKE and interaction of charges with the electric field, which is the sum of the external field and interparticle electric field. Hence let us call it the RCKE-electric field (RCKE-EF) interaction. In the self-consistent field approximation it appears as
\begin{equation}\label{SRW kin en force part}F^{\alpha}_{sr}=\frac{e\hbar^2}{4m^2c^2}\partial_\beta(\partial_\alpha
E_\beta\cdot n).\end{equation}
As the RCKE-EF interaction has semi-relativistic
origin, we can write $G_{ij}$ instead of $\tilde{G}_{ij}$ in this
term. The RCKE also gives other terms in the force field, all of
them are presented below in the Euler equation.

The force caused by the Darwin term was accounted in Ref. \cite{Asenjo arxiv 2011}, when the force field presented by formula (\ref{SRW kin en force part}) was not considered in Ref. \cite{Asenjo arxiv 2011} at derivation of the kinetic equation and calculation of the Langmuir wave dispersion. The interparticle Darwin interaction was not considered in Ref. \cite{Asenjo arxiv 2011} either. Thus, we are going to generalize result of Ref. \cite{Asenjo arxiv 2011} calculating contribution of the force field (\ref{SRW kin en force part}) in the Langmuir wave dispersion. We also present equations of semi-relativistic collective motion of plasma. But we consider hydrodynamic equations, when the kinetic equation was considered in Ref. \cite{Asenjo arxiv 2011}.

Equations (\ref{SRW Coulomb wide field}) and (\ref{SRW kin en
force part}) are very similar, but they have two differences. The
first difference is distinction in tensor structure and the second
one is the fact that equation (\ref{SRW kin en force part})
contains concentration under the spatial derivative, whereas in
formula (\ref{SRW Coulomb wide field}) concentration contains as
an external multiplier.

\section{Equations of quantum hydrodynamics}

In the previous section we have shown similarity of the RCKE-EF and the Darwin interaction. One of the aims of the paper is to compare contribution of these terms in the QHD equations and the Langmuir wave dispersion. We want to trace separate contribution of the each term. Thus we need to mark them.

The first equation of the QHD set is the continuity equation
\begin{equation}\label{SRW continuity equation}\partial_{t}n+\nabla\textbf{j}=0.\end{equation}
In that equation a function of current
$\textbf{j}(\textbf{r},t)=n(\textbf{r},t)\textbf{v}(\textbf{r},t)$
is arisen, where $\textbf{v}(\textbf{r},t)$ is the velocity
field.

The second equation of the QHD set is the Euler equation, but in the semi-relativistic approximation function $\textbf{j}(\textbf{r},t)$ appeared in the continuity equation is the particle current. However, in contrast to non-relativistic case we can not call it momentum density, thus the Euler equation is the equation of particle current evolution \cite{Andreev arxiv rel}. This equation has form
$$mn(\partial_{t}+v^{\beta}\nabla^{\beta})v^{\alpha}+\partial_\beta P_{\alpha\beta}=enE^{\alpha}+\frac{e}{c}\varepsilon^{\alpha\beta\gamma}nv^{\beta}B^{\gamma}$$
$$+\frac{e\hbar^{2}}{8m^{2}c^{2}}n\partial^{\alpha}(\partial^{\beta}E^{\beta}_{ext})
+\frac{e\hbar^{2}}{4m^{2}c^{2}}n\partial^{\alpha}(\partial^{\beta}E^{\beta}_{int})$$
$$+\frac{e\hbar^2}{4m^2c^2}\partial_\beta(\partial_\alpha
E_\beta\cdot n)$$
$$ -\frac{e}{mc^2}\biggl[E_\beta(mnv_\alpha
v_\beta+P_{\alpha\beta})+E_\alpha\biggl(\frac{1}{2}mnv^2+n\varepsilon\biggr)\biggr]$$
$$-\frac{e^2\hbar^2}{8m^2c^2}\partial_\beta n\int
d\mathbf{r}'\partial_\alpha
G_{\beta\gamma}(\mathbf{r}-\mathbf{r}')\partial^\prime_\gamma
n(\mathbf{r}',t)$$
$$-\frac{e^3}{2mc^2}n\int
d\mathbf{r}'G_{\alpha\beta}(\mathbf{r}-\mathbf{r}')E_\beta(\mathbf{r}',t)n(\mathbf{r}',t)$$
$$+\frac{e^2}{2c^2}\int d\mathbf{r}'[\partial_\alpha G_{\beta\gamma}(\mathbf{r}-\mathbf{r}')-\partial_\beta
G_{\alpha\gamma}(\mathbf{r}-\mathbf{r}')]\pi_{\beta\gamma}(\mathbf{r},\mathbf{r}',t)$$
$$+\frac{e^2}{2mc^2}n\int d\mathbf{r}'\partial_\gamma
G_{\alpha\beta}(\mathbf{r}-\mathbf{r}')\times$$
\begin{equation}\label{SRW Euler eq} \times[mn(\mathbf{r}',t)v_\beta(\mathbf{r}',t)
v_\gamma(\mathbf{r}',t)+P_{\beta\gamma}(\mathbf{r}',t)],
\end{equation}
where $\textbf{E}=\textbf{E}_{ext}+\textbf{E}_{int}$ and $\textbf{B}$ are the electric and magnetic
fields, $n\varepsilon$ is the density of thermal energy including
quantum part (which is an analog of the quantum Bohm potential),
$\varepsilon^{\alpha\beta\gamma}$ is the antisymmetric symbol
(the Levi-Civita symbol), $P^{\alpha\beta}$ is the pressure tensor, which is the semi-relativistic generalization of the sum
of non-relativistic thermal pressure $p^{\alpha\beta}$ and the
quantum Bohm potential $T^{\alpha\beta}$. In right-hand side of
equation (\ref{SRW Euler eq}) a force field locates. The force
field consists of the Lorentz force and specific quantum
semi-relativistic terms, which are discussed below.
$\pi_{\alpha\beta}(\mathbf{r},\mathbf{r}',t)$ is presented
explicitly and considered below after analysis of
$P^{\alpha\beta}$ structure. The vector potential appears
in the form
\begin{equation}\label{SRW } A^{int}_\alpha(\mathbf{r},t)=\frac{e}{2c}\int d\mathbf{r}'G_{\alpha\beta}(\textbf{r}-\textbf{r}')n(\textbf{r}',t)v_\beta(\textbf{r}',t),
\end{equation}
which gives contribution in the Lorentz force, the second term in right-hand side of equation (\ref{SRW Euler eq}) along with external magnetic field. Magnetic field $\textbf{B}=\nabla\times \textbf{A}^{int}$ satisfies to the quasi-magnetostatic Maxwell equation:
\begin{equation}\label{SRW rot B}\nabla\times \textbf{B}=\frac{4\pi}{c}\textbf{j},\end{equation}
and
\begin{equation}\label{SRW div B}\nabla\textbf{B}=0.\end{equation}

We believe that it is worthwhile to admit that we have not neglected time derivatives in the Maxwell equations (\ref{SRW Max rot}) and (\ref{SRW rot B}). We do not present these terms, because they do not appear in the semi-relativistic approximation. The Hamiltonian (\ref{SRW Hamiltonian}) contains the Coulomb interaction and the current-current interaction (the Biot--Savart law). Obtained Maxwell equations correspond to the Hamiltonian. We can put these well-known time derivatives back in the Maxwell equations. However, this step breaks logic of semi-relativistic description. So, we keep to work with the electromagnetic fields appearing in the semi-relativistic approximation and describing by equations (\ref{SRW Max rot}), (\ref{SRW Max div one species}), (\ref{SRW rot B}) and (\ref{SRW div B}).

Before discussion of pressure tensor $P^{\alpha\beta}$ we explain the physical meaning of the force field terms presented
in the right-hand side of the Euler equation (\ref{SRW Euler eq}).
It is especially important as some of these terms are presented for
the first time.

The first two terms present the density of
the Lorentz force. The self-consistent part of the Coulomb
interaction gives contribution in the first term. The second term
contains contribution of the current-current interaction in the
self-consistent field approximation. We should admit that
all terms in the Euler equation are presented in the
self-consistent field approximation. Actually, only a part of the
whole contribution from the current-current interaction came in the
Lorentz force, it also leads to several other terms. They are
seventh--tenth terms of the Euler equation. In fact, the terms
eight--ten already appear in the classic semi-relativistic
hydrodynamics, but in the quantum theory
these terms have more rich structure. First of all, they contain
contribution of the exchange interactions via quantum
correlations, which is not considered in this paper, but they
naturally appear in the many-particle QHD. We neglect them
here considering the self-consistent field approximation only. However they contain contribution of the quantum Bohm potential along with the thermal pressure.

The third term corresponds to the Darwin term. The fourth term corresponds to the Darwin interaction. Terms three and four have same nature, but they have different coefficients. The Darwin term appears from the Dirac equation describing motion of an electron in an external field. Then the Darwin interaction comes from quantum electrodynamic scattering amplitude of two particles. The third and fourth terms contain electric field. The third (fourth) term includes the external (interparticle Coulomb) electric field $\textbf{E}_{ext}$ ($\textbf{E}_{int}$). Due to different coefficients before these terms we can not combine them together having full electric field $\textbf{E}=\textbf{E}_{ext}+\textbf{E}_{int}$. We should not expect additivity of electric fields in these terms, since formula (\ref{SRW Coulomb wide field}) used in the fourth term is asymptotic. Original force field in presented by formula (\ref{SRW Coulomb wide concentr}). The original formula does not contain electromagnetic field. It is presented in terms of the particle concentration of interacting species $\textbf{F}_{D}=\frac{\pi e^{2}\hbar^{2}}{m^{2}c^{2}}n\nabla n$. In our case we have one species, so it describes an electron-electron interaction. In this paper we consider plasmas without external electric field. Consequently the third term equals to zero. In Ref. \cite{Asenjo arxiv 2011} authors have deal with the Darwin term appearing from the Dirac equation, which corresponds to the interaction of electrons with the external field. As a consequence they get coefficient in two times less then we get from the interparticle interaction.

The terms five and six present contribution of the RCKE. The fifth
term has simple structure, it contains divergence $\nabla^{\beta}$ of
the tensor which is product of particle concentration on $\nabla^{\alpha}E^{\beta}$ and presents the RCKE-EF interaction.
In the sixth term, which contains a number of terms in square brackets, the first set of them is the convolution of $E^{\beta}$ with the tensor which is the current of the particle current $\textbf{j}$, and as a part of this current we have the pressure tensor $P^{\alpha\beta}$. As the sixth term of the Euler equation has semi-relativistic nature, we should consider only non-relativistic part of $P^{\alpha\beta}$. The second set of terms in the sixth term is the product of electric field $E^{\alpha}$ on the energy density. The energy density was separated on two parts there. First of them is the kinetic energy density of a local ordered motion.  We need to say that $n\varepsilon$ is the energy density which consists of two parts: thermal energy and quantum contribution -- an analog of the quantum Bohm potential. In one particle case we lose contribution of thermal motion and quantum-thermal terms, and get quantum terms arising for non-interacting particles. $\varepsilon$ gives no contribution in considered below problem, therefore we do not present its explicit form.

Explicit form of the tensor $P^{\alpha\beta}$ is
$$P_{\alpha\beta}(\textbf{r},t)=\int
dR\sum_{i=1}^N\delta(\mathbf{r}-\mathbf{r}_i)a^2\times$$
$$\times\biggl[m
u_{i\alpha}u_{i\beta}-\frac{\hbar^2}{2m}\biggl(1-\frac{v_i^2}{c^2}\biggr)
\partial_{i\alpha}\partial_{i\beta}\ln a$$
$$+\frac{\hbar^2}{2m
c^2}(\partial_{i\alpha}v_{i\gamma}\partial_{i\beta}v_{i\gamma}+v_{i\gamma}
\partial_{i\alpha}\partial_{i\beta}v_{i\gamma})$$
$$+\frac{\hbar^2}{4m c^2}(v_{i\alpha}\partial_{i\beta}+v_{i\beta}\partial_{i\alpha})(\partial_{i\gamma}v_{i\gamma}+
2v_{i\gamma}\partial_{i\gamma}\ln a)$$
$$-\frac{\hbar^4}{4m^3 c^2 a^{2}}\biggl(a\partial_{i\alpha}\partial_{i\beta}\Delta_i a+\partial_{i\alpha}\partial_{i\beta}a\Delta_i a$$
$$-\partial_{i\alpha}a\partial_{i\beta}\Delta_i a-\partial_{i\beta}a\partial_{i\alpha}\Delta_i
a\biggr)\biggr]$$
$$+\int dR\sum_{i=1,j=1,i\neq
j}^N\delta(\mathbf{r}-\mathbf{r}_i)a^2\frac{\hbar^2e^2}{4m^2c^2}\times$$
\begin{equation}\label{SRW pressure general}\times(G^{\beta\gamma}_{ij}
\partial_{i\alpha}\partial_{j\gamma}\ln a+G^{\alpha\gamma}_{ij}
\partial_{i\beta}\partial_{j\gamma}\ln a),\end{equation}
where $v_{i\alpha}$ is the velocity of i-th particle, and it is the sum of the velocity field $v^{\alpha}(\textbf{r},t)$ and thermal velocity $u_{i}^{\alpha}$, $a$ is the amplitude of the wave function $\psi(R,t)=a\exp(iS/\hbar)$, velocity of i-th particle $v_{i}^{\alpha}$ connects with the phase of the wave function as
$$v_i^\alpha =\frac{s_i^\alpha}{m_i}-\frac{s_i^\alpha s_i^2}{2m_i^3 c^2}
+\frac{\hbar^2}{2m_i^3c^2}\biggl[s_i^\alpha a^{-1}\Delta_i a+\partial_i^\alpha(s_i^\beta\partial_i^\beta \ln a)$$
\begin{equation}+\frac{1}{2}\partial_i^\alpha\partial_i^\beta s_i^\beta\biggr]
-\sum_{j=1,j\neq i}^N\frac{e_i e_j}{2m_i m_j c^2}G^{\alpha\beta}_{ij}s_j^\beta ,\end{equation}
where $s_i^\alpha=\partial_i^\alpha S-\frac{e_i}{c}A_i^\alpha$. The first term in formula (\ref{SRW pressure general}) is the non-relativistic thermal pressure, the second term in this formula consist of two parts, the first of them is the non-relativistic quantum Bohm potential, the other terms present semi-relativistic effects. Neglecting thermal velocities in the semi-relativistic terms of the pressure tensor $P_{\alpha\beta}$ we get purely quantum semi-relativistic pressure, which is the semi-relativistic generalization of the quantum Bohm potential $T_{\alpha\beta}$, which explicit form for ideal gas is
\begin{equation}\label{SRW Bohm}T_{\alpha\beta}=-\frac{\hbar^{2}}{4m}\partial^{\alpha}\partial^{\beta}
n+\frac{\hbar^{2}}{4m}\Biggl(\frac{\partial^{\alpha}n\cdot\partial^{\beta}n}{n}\Biggr).\end{equation}
We also drop contribution of the current-current interaction. In
the result we have
$$P_{\alpha\beta}(\textbf{r},t)=p_{\alpha\beta}+T_{\alpha\beta}-\frac{v^{2}}{c^{2}}T_{\alpha\beta}$$
$$+\frac{\hbar^{2}}{2mc^{2}}n(\partial^{\alpha}v^{\gamma}\partial^{\beta}v^{\gamma}+v^{\gamma}\partial^{\alpha}\partial^{\beta}v^{\gamma})$$
$$+\frac{\hbar^{2}}{4mc^{2}}n(v^{\alpha}\partial^{\beta}+v^{\beta}\partial^{\alpha})(\nabla \textbf{v})$$
$$+\frac{\hbar^{2}}{4mc^{2}}(\partial^{\gamma}n)\biggl(v^{\alpha}\partial^{\beta}v^{\gamma}+v^{\beta}\partial^{\alpha}v^{\gamma}\biggr)$$
$$-\frac{1}{c^{2}}\biggl(v^{\alpha}v^{\gamma}T^{\beta\gamma}+v^{\beta}v^{\gamma}T^{\alpha\gamma}\biggr)$$
$$-\frac{\hbar^{4}}{4m^{3}c^{2}}\biggl(\sqrt{n}\cdot\partial^{\alpha}\partial^{\beta}\triangle\sqrt{n}+\partial^{\alpha}\partial^{\beta}\sqrt{n}\cdot\triangle\sqrt{n}$$
\begin{equation}\label{SRW pressure simple}-\partial^{\alpha}\sqrt{n}\cdot\partial^{\beta}\triangle\sqrt{n}-\partial^{\beta}\sqrt{n}\cdot\partial^{\alpha}\triangle\sqrt{n}\biggr).\end{equation}
The first two terms have non-relativistic nature. The other terms
are semi-relativistic, most of them are proportional to
$v^{2}/c^{2}$, except of the four last terms.
Thermal pressure $p_{\alpha\beta}$ does not depend on interaction,
so we can use equation of state for ideal gas, and we write
$p^{\alpha\beta}=nk_{B}T\delta^{\alpha\beta}$, where $k_{B}$ is
the Boltzmann constant, $T$ is the temperature,
$\delta^{\alpha\beta}$ is the Kronecker symbol. When
$P^{\alpha\beta}$ stays in a semi-relativistic term we should
neglect semi-relativistic part and consider non-relativistic one only.

Let us repeat a part of the semi-relativistic quantum Bohm potential existing in linear approximation, assuming that an equilibrium condition is described by non-zero uniform concentration $n_{0}\neq 0$ and zero velocity field $\textbf{v}_{0}=0$. Hence we obtain
\begin{equation}\label{SRW Bohm lin} \partial_{\beta}T^{\alpha\beta}_{lin}=-\frac{\hbar^{2}}{4m}\partial^{\alpha}\triangle
n -\frac{\hbar^{4}}{8m^{3}c^{2}}\partial^{\alpha}\triangle\triangle n. \end{equation}
We have two terms. The first of them appears from the first term in the formula (\ref{SRW Bohm}). The second term in formula (\ref{SRW Bohm lin}) is the linear part of the first term in the last group of terms, which does not contain the velocity field, in formula (\ref{SRW pressure simple}).

Here we present explicit form of $\pi_{\alpha\beta}(\mathbf{r},\mathbf{r}',t)$, which is the part of the seventh term in the force field
$$\pi_{\alpha\beta}(\mathbf{r},\mathbf{r}',t)=\int\prod_{j=1}^N d\mathbf{r}_j\sum_{i,j=1,i\neq j}^N\delta(\mathbf{r}-\mathbf{r}_i)\delta(\mathbf{r}'-\mathbf{r}_j)\times$$
$$\times a^2(u_{i\alpha}u_{j\beta}-
\frac{\hbar^2}{2m^2}\partial_{i\alpha}\partial_{j\beta}\ln a).$$
To close the QHD set of equations we should find approximate
connection between $\pi_{\alpha\beta}(\mathbf{r},\mathbf{r}',t)$
and other hydrodynamic quantities. Calculating
$\pi_{\alpha\beta}(\mathbf{r},\mathbf{r}',t)$ for the system of
independent particles we get
$\pi_{\alpha\beta}(\mathbf{r},\mathbf{r}',t)=0$. Thus, in the
first approximation we do not need to account contribution of
$\pi_{\alpha\beta}(\mathbf{r},\mathbf{r}',t)$ in the QHD
equations.

\section{Dispersion equation for quantum semi-relativistic Langmuir waves}

To get semi-relativistic effects in the form of analytic simple
formulas we consider quantum motion of electrons on the background of motionless ions.
We consider the small perturbations of equilibrium state like
$$\begin{array}{ccc}n=n_{0}+\delta n, & \textbf{v}=0+\delta\textbf{v}  \end{array},$$
$$\begin{array}{ccc}\textbf{E}=0+\delta\textbf{E},&
\textbf{B}=0+\delta \textbf{B},& \delta p=3mv_{s}^{2}\delta n\end{array},$$
where $m$ is the mass of the electron. In equations (\ref{SRW continuity equation}), (\ref{SRW Euler eq}) and the Maxwell equations (\ref{SRW Max rot}), (\ref{SRW Max div}), (\ref{SRW rot B}) and (\ref{SRW div B}), $v_{s}^{2}$ is the average thermal velocity (for the case
of degenerate electrons we should write $\frac{v_{Fe}^{2}}{3}(1-\frac{1}{10}\frac{v_{Fe}^{2}}{c^{2}})$, instead of $3v_{s}^{2}$, see Appendix, where $v_{Fe}=\sqrt[3]{3\pi^{2}n_{0}}\hbar/m$ is the Fermi velocity).
Substituting these relations into the set of equations and neglecting by
nonlinear terms, we obtain a system of linear homogeneous
equations in partial derivatives with constant coefficients.
Carrying the following representation for small perturbations
$\delta f$
$$\delta f =f(\omega, \textbf{k}) exp(-\imath\omega t+\imath \textbf{k}\textbf{r}) $$
yields a homogeneous system of algebraic equations.

The Euler equation (\ref{SRW Euler eq}) is very complicated, thus we allow ourselves to present the algebraic form of linearized Euler equation
$$-\imath\omega mn_{0} \delta v^{\alpha}+\imath k^{\alpha} \biggl(3mv_{se}^{2} +\frac{\hbar^{2}k^{2}}{4m}-\frac{\hbar^{4}k^{4}}{8m^{3}c^{2}}\biggr)\delta n$$
\begin{equation}\label{SRW lin Euler}=en_{0}\delta E^{\alpha} -k^{\alpha}k^{\beta}2\frac{en_{0}\hbar^{2}}{4m^{2}c^{2}}\delta E^{\beta}-\frac{e^{3}n_{0}^{2}}{2mc^{2}}G^{\alpha\beta}(\textbf{k})\delta E^{\beta},\end{equation}
where $G^{\alpha\beta}(\textbf{k})$ is the Fourier image of the current-current interaction Green function (\ref{SRW CC Gr func}), its explicit form is
$$G^{\alpha\beta}(\textbf{k})=\frac{8\pi}{k^{2}}\biggl(\delta^{\alpha\beta}-\frac{k^{\alpha}k^{\beta}}{k^{2}}\biggr).$$
The last term in the Euler equation (\ref{SRW Euler eq}) gives a
linear term due to the linear part of the quantum Bohm potential,
which is a part of $P^{\beta\gamma}$, but it is equal to zero
because of the structure of $G^{\alpha\beta}(\textbf{k})$.
The last term in equation (\ref{SRW lin Euler}) gives no
contribution in the dispersion of the Langmuir waves. We also admit that we do not consider the temperature-relativistic effects $\sim T/mc^{2}$.

The electric field is assumed to have a nonzero value.
Expressing all quantities through the electric field, we come to the equation
$$\omega^{2}=\omega_{Le}^{2}\biggl(1-\frac{\hbar^{2}k^{2}}{2m^{2}c^{2}}\biggr)$$
\begin{equation}\label{SRW general form of disp eq}
+\Biggl(\left(
                     \begin{array}{c}
                       3v_{se}^{2} \\
                       \frac{1}{3}v_{Fe}^{2}(1-\frac{1}{10}\frac{v_{Fe}^{2}}{c^{2}}) \\
                     \end{array}
                   \right)
+\frac{\hbar^{2}k^{2}}{4m^{2}}-\frac{\hbar^{4}k^{4}}{8m^{4}c^{2}}\Biggr)k^{2}, \end{equation}
where $\omega_{Le}$ is the Langmuir frequency, $\omega_{Le}^{2}=4\pi e^{2}n_{0}/m$. The first group of terms in the right-hand side of (\ref{SRW general form of disp eq}) consists of three parts: the Langmuir frequency, contribution of the RCKE-EF and Darwin interactions, where the RCKE-EF interaction leads to $-\frac{\hbar^{2}k^{2}\omega_{Le}^{2}}{4m^{2}c^{2}}$, and the Darwin interaction leads to the same structure $-\frac{\hbar^{2}k^{2}\omega_{Le}^{2}}{4m^{2}c^{2}}$. Together they give $-\frac{\hbar^{2}k^{2}\omega_{Le}^{2}}{2m^{2}c^{2}}$. The second group in equation (\ref{SRW general form of disp eq}) consists of three parts: contribution of the pressure (thermal motion or Fermi pressure), the next part is the well-known quantum Bohm potential, the last part is the contribution of the RCKE via the semi-relativistic part of the pressure tensor, or, speaking in other terms, it is the semi-relativistic part of the quantum Bohm potential.

Comparing formula (\ref{SRW general form of disp eq}) with the results of Ref. \cite{Asenjo arxiv 2011} we should mention several differences.
In Ref. \cite{Asenjo arxiv 2011} authors do not have the quantum Bohm potential contribution $\sim\hbar^{2}k^{4}$. They also do not obtain the semi-relativistic part of the quantum Bohm potential. Most dramatic difference arises at consideration of the second term in the first group of terms in formula (\ref{SRW general form of disp eq}) with the corresponding result of Ref. \cite{Asenjo arxiv 2011}. Here we have $-\frac{1}{8}\frac{\hbar^{2}k^{2}\omega_{Le}^{2}}{m^{2}c^{2}}$. In Ref. \cite{Asenjo arxiv 2011} it was found as $+\frac{1}{2}\frac{\hbar^{2}k^{2}\omega_{Le}^{2}}{m^{2}c^{2}}$. We have difference in signs and magnitude of coefficients. It looks like they choose another sign before the Darwin term (see Ref. \cite{Asenjo arxiv 2011} the fifth term in formula (5) and formula (\ref{SRW Hamiltonian Darwin}) of our paper). As we mention they neglected the RCKE $\textbf{D}^{4}$, because they did not want to consider the semi-relativistic part of the quantum Bohm potential. As we have shown the RCKE gives several different contributions. Consequently the RCKE-EF interaction was also lost in Ref. \cite{Asenjo arxiv 2011}, which gives half of the term under consideration. Moreover they applied, for the electron-electron interaction, the Darwin term describing interaction of the charges with external electric field. Which is two times smaller than the electron-electron Darwin interaction (see terms three and four in the right-hand side of formula (\ref{SRW Euler eq}) of our paper). Altogether we see why the coefficient obtained in Ref. \cite{Asenjo arxiv 2011} in four times smaller than our results.

\subsection{Estimations}

In this subsection we consider system of degenerate electrons. Our aim is to find parameters of system when semi-relativistic effects are noticeable. To this end we represent spectrum of the semi-relativistic Langmuir waves in terms of the Bohm velocity $v_{b}$ defined as $v_{b}^{2}\equiv\hbar^{2}k^{2}/(4m^{2})$. The Bohm velocity is not a constant since it depends on the wave vector $v_{b}=v_{b}(k)$. The spectrum reappears in the following form
$$\omega^{2}=\omega_{Le}^{2}\biggl(1-2\frac{v_{B}^{2}}{c^{2}}\biggr)$$
\begin{equation}\label{SRW general form of disp eq for Estim} +\frac{1}{3}v_{Fe}^{2}\biggl(1-\frac{1}{10}\frac{v_{Fe}^{2}}{c^{2}}\biggr)k^{2} +v_{B}^{2}k^{2}\biggl(1-2\frac{v_{B}^{2}}{c^{2}}\biggr). \end{equation}
Contribution of the semi-relativistic effects is noticeable at large Bohm velocity. The Bohm velocity $v_{b}$ increased with increasing of the wave vector, which is bounded above. This limitation is related to average interparticle distance $a$ giving minimal wavelength, and, consequently, maximal wave vector $k_{max}\sim\frac{1}{a}\sim\sqrt[3]{n_{0}}$. For parameters included in the spectrum (\ref{SRW general form of disp eq for Estim}), at small wavelength limit, we find $\omega_{Le}^{2}\sim n_{0}$, $v_{Fe}^{2}\sim n_{0}^{2/3}$, $k^{2}\sim\frac{1}{a^{2}}\sim n_{0}^{2/3}$, $v_{B}^{2}\sim k^{2}\sim n_{0}^{2/3}$. We are interested in a regime when the Bohm velocity $v_{B}$ is comparable with the speed of light $c$, but $v_{B}\ll c$. Otherwise we do not get the semi-relativistic regime. Let us estimate parameters of system at $\frac{v_{B}^{2}}{c^{2}}\approx0.1$. In this case the semi-relativistic effects RCKE-EF+Darwin interactions and the semi-relativistic quantum Bohm potential decrease the corresponding terms on two percent. We obtain $k_{max}=1.2$ $10^{10}$ cm$^{-1}$. It corresponds to the particle concentration $n_{0}\approx10^{30}$ cm$^{-3}$. Such huge densities corresponds astrophysical objects, like white dwarfs and atmosphere of neutron stars.

Similar to (\ref{SRW general form of disp eq for Estim}) solutions were found in Refs. \cite{Zhu PPCF 12}, \cite{Mendonca PP 11} and reviewed in Ref. \cite{Uzdensky arxiv review 14}, where other methods of plasma description were applied. Let us mention that solutions have similar structure, but instead of the Bohm velocities they have the Fermi velocities (see Ref. \cite{Uzdensky arxiv review 14} formula 47).

\section{Conclusion}

We gave derivation of the many-particle QHD equations for the
semi-relativistic system of spinless charged particles. Contribution of the RCKE, Coulomb, Darwin and current-current
interactions in the Euler equation is obtained. Contributions from different terms are compared. It is shown that
simultaneous account of the RCKE and Darwin interaction is
necessary, because the RCKE gives a number of terms having
different structure, and one of them has structure of the term
connected with the Darwin interaction in the Hamiltonian. The RCKE leads to the complex
structure of the pressure tensor. The semi-relativistic part of
the pressure tensor contains terms proportional to the ratio of the
velocity field to the square of light speed. It also includes the
several terms proportional to $\hbar^{4}/c^{2}$ and contains more
higher spatial derivatives than in the non-relativistic quantum
Bohm pressure.

Using developed approximation of the QHD equations we studied
dispersion dependence $\omega(\textbf{k})$ of the
semi-relativistic Langmuir waves. We got contribution of the RCKE,
which gives two terms in $\omega(\textbf{k})$, they are the RCKE-EF interaction and the semi-relativistic part of the quantum Bohm potential contributions, and the Darwin giving one term. We have obtained that the RCKE-EF and the Darwin interactions gives an equal contribution in the dispersion dependence of the Langmuir waves.

We have developed the QHD method with all semi-relativistic effects in spinless plasmas for further research of linear and nonlinear effects in the semi-relativistic quantum plasmas.

The spinless semi-relativistic effects play important role in plasmas of spinless particles and in plasmas of spinning particles. Spin-dependent semi-relativistic interactions were considered in literature earlier: the spin-spin interaction was studied in Refs. \cite{MaksimovTMP 2001}, \cite{Andreev RPJ 07} \cite{Andreev arxiv MM}, the spin-current interaction is considered in \cite{Andreev RPJ 07}, the spin-orbit interaction was included in the QHD equations in Refs. \cite{Andreev arxiv MM} and \cite{pavelproc}.

We conclude that this paper fulfils the programm of development of the semi-relativistic hydrodynamics based on the Breit Hamiltonian.

\section{Appendix: Equation of state for semi-relativistic degenerate Fermi gas}

For relativistic fermions, assuming relativistic relation between energy and momentum of each particle $\epsilon_{i}=\sqrt{p_{i}^{2}c^{2}+m^{2}c^{4}}$ and applying usual technics (see \cite{Landau v5} see sections 56, 58), one can find equation of state
\begin{equation}\label{SRW rel eq of state} p=\frac{\pi}{3}\frac{m^{4}c^{5}}{(2\pi\hbar)^{3}}\Xi\biggl(\frac{p_{F}}{mc}\biggr),\end{equation}
where $p_{F}=(3\pi^{2}n_{0})^{1/3}\hbar$ is the Fermi momentum, and
\begin{equation}\label{SRW rel eq of state} \Xi(x)=8\int_{0}^{x}\frac{\xi^{4}}{\sqrt{1+\xi^{2}}}d\xi.\end{equation}

We consider the semi-relativistic limit. Consequently we have $\frac{p_{F}}{mc}\ll1$, and $x\ll1$, $\xi\ll1$.

In the semi-relativistic approximation we can write equation of state in the following form
\begin{equation}\label{SRW rel eq of state Semi-rel} p=p_{NR}\biggl(1-\frac{1}{14}\frac{v_{Fe}^{2}}{c^{2}}\biggr),\end{equation}
where we have used $p_{NR}$ for the non-relativistic Fermi pressure $p_{NR}=\frac{\hbar^{2}}{5m}(3\pi^{2})^{\frac{2}{3}}n^{\frac{5}{3}}$.

For linear perturbation of the pressure (\ref{SRW rel eq of state Semi-rel}) we obtain
\begin{equation}\label{SRW rel eq of state Semi-rel perturbation} \delta p=\frac{\partial p}{\partial n}\delta n=\frac{1}{3}mv_{Fe}^{2}\biggl(1-\frac{1}{10}\frac{v_{Fe}^{2}}{c^{2}}\biggr)\end{equation}

\end{document}